\documentclass[fleqn,11pt]{article}

\usepackage{here}
\usepackage{amsmath, amsthm, amsfonts}
\RequirePackage{times}      
\RequirePackage{graphicx,xcolor}
\RequirePackage[left=2cm,%
                right=2cm,%
                top=2.25cm,%
                bottom=2.25cm,%
                headheight=12pt,%
                letterpaper]{geometry}%

\title{Completing density functional theory by machine learning hidden messages from molecules}
\author{Ryo Nagai$^{1,2,*}$, Ryosuke Akashi$^{1}$, Osamu Sugino$^{1,2}$\\
\small ${}^1$Department of Physics, The University of Tokyo, Hongo, Bunkyo-ku, Tokyo 113-0033, Japan\\
\small ${}^2$Institute for Solid State Physics, The University of Tokyo, Kashiwa, Chiba 277-8581, Japan\\
\small ${}^*$+81-47-136-3292, nagai-ryo236@g.ecc.u-tokyo.ac.jp
}

\begin{document}
\maketitle


\flushbottom

\thispagestyle{empty}

\begin{abstract}

Kohn-Sham density functional theory is the base of modern computational approaches to electronic structures. 
Their accuracy vitally relies on the exchange-correlation energy functional, which encapsulates electron-electron interaction beyond the classical one. The functional provides a way to obtain the density and energy without solving the many-body equation and can, in principle, be determined to reproduce the exact ones universally. However, the past approaches are dependent on the theoretical development, which limits the possibility of the functional to human's intuition. 
Here, we demonstrate a systematic way to machine-learn a functional from a database, without complicated assumptions. The density and energy are related with a flexible feed-forward neural network, which is trained to reproduce accurate dataset, and the KS-DFT is solved by taking the functional derivatives with the back-propagation technique. 
Surprisingly, a trial functional, trained for just a few molecules, has been shown to be applicable to hundreds of molecular systems with comparable accuracy to the standard functionals.
Also, by adding the nodes connected to the hidden layers, a non-local term is straightforwardly included to improve accuracy, which has been hitherto impractically difficult.
Utilizing the strategy of rapidly advancing machine learning techniques, this novel approach is expected to enrich the DFT framework by constructing a functional just from a database for materials conventionally difficult to calculate accurately.

\end{abstract}

\section*{Introduction}

Kohn-Sham (KS) density functional theory (DFT)~\cite{HK, KS} is regarded as the standard method in theoretical study on electronic structures of materials. Although its practical accuracy has been established through applications to a wide range of systems including molecular and bulk materials, there is a growing need for more accurate and universal functional to extend the scope of its application.
In this theory, the solution of the KS equation
\begin{eqnarray}
\!\!\!\!\!\!\!\!\!
\left[
-\frac{\nabla^2}{2}
\!+\!
V_{\rm ion}({\bf r})
\!+\!\!\!
\int \! \! \! d{\bf r}' \!
\frac{n({\bf r}')}{|{\bf r}\!-\!{\bf r}'|}
\!+\!
V_{\rm xc}({\bf r})
\right] \! \varphi_{i}({\bf r})
\!=\!
\varepsilon_{i} \varphi_{i}({\bf r}),
\label{eq:KS}
\end{eqnarray}
with the density $n({\bf r})$ calculated by summing $|\varphi_{i}({\bf r})|^2$ for all the occupied states, yields the total energy as well as electron density distribution of an interacting electron system under the ionic potential $V_{\rm ion}$. The exchange-correlation (xc) potential $V_{\rm xc}$ is ideally a non-local functional, where its value at ${\bf r}$ is affected by the whole density distribution\{$n$\}. However, its explicit form has yet been elusive. Enormous effort has been devoted to development of functionals either in purely local forms, where $V_{\rm xc}({\bf r})$ depends only on $n({\bf r})$, or in semi-locally corrected forms using density gradient $\nabla n({\bf r})$ or KS orbital $\{\varphi_{i}\}$, thereby gradually climbing the so-called Jacob's ladder~\cite{Jacob} of accuracy at the price of complexity and computational cost. In order to determine the actual functional form, complicated and heuristic procedures have been required: One generally has to exploit elaborate approximations based on the perturbation theory, analyses of asymptotic behaviors of physical quantities, and fitting to references such as numerically expensive calculations and experimental observations. Moreover, Medvedev and coworkers have warned that, although some modern functionals thus formulated could yield highly accurate energy-related quantities, they give less accurate density distribution than those from previous functionals~\cite{Medvedev49}.

The pivotal problem is that, unlike the perturbation theory for the physical quantities~\cite{manybody}, the formula of $V_{\rm xc}$ is difficult to derive since it is defined implicitly -- the function that reproduces the exact \{$n$\} through Equation~(\ref{eq:KS}) in any systems. Alternatively, a machine-learning (ML) approach was proposed to construct an explicit form of $V_{\rm xc}$ \cite{NNVHxc, MLxc2}. 
This idea was tested previously for a one-dimensional (1D) model by training the neutral-network (NN) mapping $n\rightarrow V_{\rm xc}$ to reproduce accurate reference pairs of [$n$, $V_{\rm xc}$], which had been prepared by numerically solving the many-body Schr\"odinger equation under various external potentials \cite{NNVHxc, MLxc2}. The resulting NN xc-potential was demonstrated to be applicable (transferable) even to unreferenced external potentials. 
Without any physical conditions for the xc-potential in that model system, the ML scheme remarkably extracted the essential structure of $V_{\rm xc}$.  This result suggests a possibility to accurately construct  $V_{\rm xc}$ for a system where the asymptotic behavior or theoretical guide is not available.
 We demonstrate this is also the case for real three-dimensional and spin-polarized materials, and suggest a systematic improvement of a functional by introducing complexity over the conventional theoretical construction, deriving from a database.

\begin{figure*}[p!]
  \begin{center}
      \includegraphics[clip,width=8.6cm]{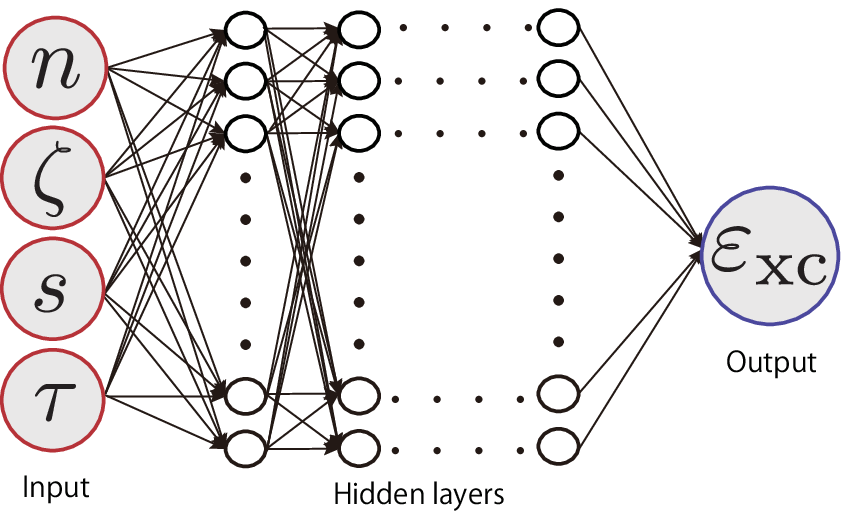}
      \caption{Structure of NN-based meta-GGA type functional. }
      \label{fig:NNmGGA}
  \end{center}
\end{figure*}

The previous studies take the whole density distribution as the input of the ML with fixing the size of target systems. When applying this NN method to real materials, however, it is impractical as the size of target system can be changed. This problem can be avoided by splitting the xc-energy into local energy density, as most conventional functionals do.
Precisely, we assume the following form for the xc-energy $E_{\rm xc}[n]$ to obtain the xc-potential by $V_{\rm xc}({\bf r})=\delta E_{\rm xc}/\delta n({\bf r})$:
\begin{eqnarray}
    && E_{\rm xc}[ n]\equiv \int {d{\bf r} ~ n ({\bf r})\varepsilon_{\rm xc}}(\mbox{\boldmath $g$}[ n] ({\bf r})),
    \label{eq:xcenergy}
\end{eqnarray}
where $\mbox{\boldmath $g$}[ n] ({\bf r})$ represents any local or non-local variables (descriptors) to include effect of the density distribution around ${\bf r}$. Most of the existing functionals adopt local spin density approximation(LSDA)~\cite{Slater,VWN}, generalized gradient approximation(GGA)~\cite{Becke-exchange, LYP, PBE}, or meta-GGA~\cite{TPSS,SCAN,M06L}, by defining $\mbox{\boldmath $g$}[ n] ({\bf r})$ as 
$(  n ({\bf r}),\zeta ({\bf r})\equiv ( n_\uparrow  ({\bf r})- n_\downarrow ({\bf r}))/ n({\bf r}))$, 
$(  n ({\bf r}),\zeta ({\bf r}),s ({\bf r})\equiv |\nabla  n ({\bf r})|/ [2(3\pi^2)^{1/3}n^{4/3}({\bf r})])$, 
or $(  n ({\bf r}),\zeta ({\bf r}), s ({\bf r}), \tau ({\bf r})\equiv 1/2 \sum_i^{\rm occ} |\nabla \phi_i ({\bf r})|^2 )$ 
respectively. In this study, the xc-energy density $\varepsilon_{\rm xc}({\bf r})$ is formulated with the feed-forward NN with $H$ layers, which is a vector-to-vector mapping ${\bf u}\rightarrow {\bf v}$ represented by
\begin{eqnarray}
   {\bf v}=h_H(...(h_2(h_1({\bf u})),
    \label{eq:NN structure}\\
     h_{i}({\bf x}) \equiv f({\bf x}W_i+{\bf b}_{i}).
     \label{eq:NN layer}
\end{eqnarray}
where the $h_i$ represents the $i$-th layer of the NN, and the input vector $\bf x$ is non-linearly transformed by the activation function $f$ after linearly transformed by the weight parameters $W_i$ and ${\bf b}_i$. 
To evaluate the functional derivative of $\delta E_{\rm xc}/\delta n({\bf r})$, we utilize the backpropagation technique~\cite{backpropagation}, which is an efficient algorithm to differentiate a NN applying the chain rule. This NN form thus relates $\{n({\bf r})\}$ and $\{V_{\rm xc}({\bf r})\}$ to be incorporated into the KS equation. In this case, we define ${\bf u}$ as the local density descriptors $\mbox{\boldmath $g$}[ n] ({\bf r})$ and ${\bf v}$ as one-dimensional vector $\varepsilon_{\rm xc}({\bf r})$ (Fig.~\ref{fig:NNmGGA}). See the Method section for more details.

There exist some remarkable advantages of incorporating the NN scheme into KS-DFT. The intrinsically complicated structure of the xc-functional can be reproduced by the NN because of its ability to represent any well-behaved functions just by increasing the number of nodes~\cite{HORNIK1991251}, which in our case is the dimension of the matrices $W_1$ to $W_H$. By optimizing (training) its parameters in the spirit of ML -- extracting hidden relationship from a database without assumption, we expect that the NN can extract the essence of electron-electron interaction and systematically imitate the complicated xc-functional. ML approaches have been previously applied to relate structural properties of materials with physical quantities with bypassing an electronic structure theory, as is done in the materials informatics~\cite{crystal_graph,Chandrasekaran2019}, but the trained ML model thus obtained often fails to be applicable for materials whose constituent elements are not covered in the training dataset. On the contrary, by taking the electron density only as the input of ML as is done in the orbital free DFT~\cite{Snyder,Burke-bypassing}, one can expect better transferability because the density is a common property of material.
The orbital free DFT, however, requires us to construct the functional for the kinetic energy of an interacting system, which should be more challenging than to explicitly evaluating the non-interacting kinetic energy in the KS-DFT framework. Also, within the KS-DFT framework, incorrect spatial oscillation in the density distribution caused by an overfitting can be suppressed as shown in ref.\cite{NNVHxc}.
For those reasons, we target the construction of xc-functional expecting that the the electrons' property can be learned most easily, with reducing the prerequisite cost of preparing the training dataset and leading to higher transferability, as has the conventional KS-DFT functionals been successful for various materials.

We prepared four types of input vector $\mbox{\boldmath $g$}$: LSDA, GGA, meta-GGA, and also a new formulation that we call ``near region approximation'' (NRA) by defining $\mbox{\boldmath $g$}[ n] ({\bf r})=\left(  n ({\bf r}),\zeta ({\bf r}), s ({\bf r}), \tau ({\bf r}), R({\bf r}) \right)$, \\where $R({\bf r}) \equiv \int  d{\bf r'} n({\bf r'}) \exp (-|{\bf r- {\bf r'}}|/\sigma)$. Gunnarsson et al.~\cite{AD} showed that such an averaged density around ${\bf r}$ describes $\varepsilon_{\rm xc}(\bf r)$ efficiently, therefore we added it into the $\mbox{\boldmath $g$}$ of meta-GGA. Construction of this non-local functional has not been popular except for the van der Waals functional, because of (a) absence of appropriate asymptotic form or other theoretical guide and (b) difficulty of fitting.

To test the performance of this method, we constructed a functional using only a few molecules with their optimized geometry to train the NN.
We chose the three molecules according to the following criteria: The structures of the molecules should be distinct from each other and have low symmetry. Electrically-polarized molecules are preferred to include to deal with cases where optimized orbitals are highly distorted from the atomic orbitals. 
Also, it is most important to include at least one spin-polarized molecule, which is necessary for determining the dependency on spin-polarization $\zeta$. Following those criteria, three simple molecules, H$_2$O, NH$_3$, and NO are selected as the the reference molecules. Note that the NO radical is spin-polarized.

The functionals are trained to reproduce the atomization energy (AE) and the density distribution (DD) of the reference molecules. We prepared the energy and density by performing accurate quantum chemical calculations~\cite{CCSD,G2H0}, which are more accurate methods than DFT.
 We choose AE instead of the total energy (TE) considering that typical errors by conventional functionals for TE ($\sim$hartree) is much larger than that for AE ($\sim$eV or kcal/mol. See Table 1.). The larger error would suggest the difficulty of reproducing TE within the semi-local approximations, while the relative energy such as AE can be predicted more accurately due to the error cancellation. It is also worth emphasizing that DD contains abundant information of electronic structure all over the three-dimensional space, which can contribute to determining the large number of NN parameters. We choose the above conditions just for demonstration, although it remains a target of future study how the accuracy depends on the choice of training dataset.

The training was performed by a Metropolis-type update of NN parameters to reproduce the training data: At each Monte-Carlo step, KS-DFT calculation is performed using the NN-based functional under training, and errors for AE and DD of the three molecules are evaluated for update of parameters. This procedure is repeated until the errors become minimized. See the Method section for computational details.

\section*{Results and Discussions}

\begin{figure*}[h]
  \begin{center}
      \includegraphics[clip,width=17.8cm]{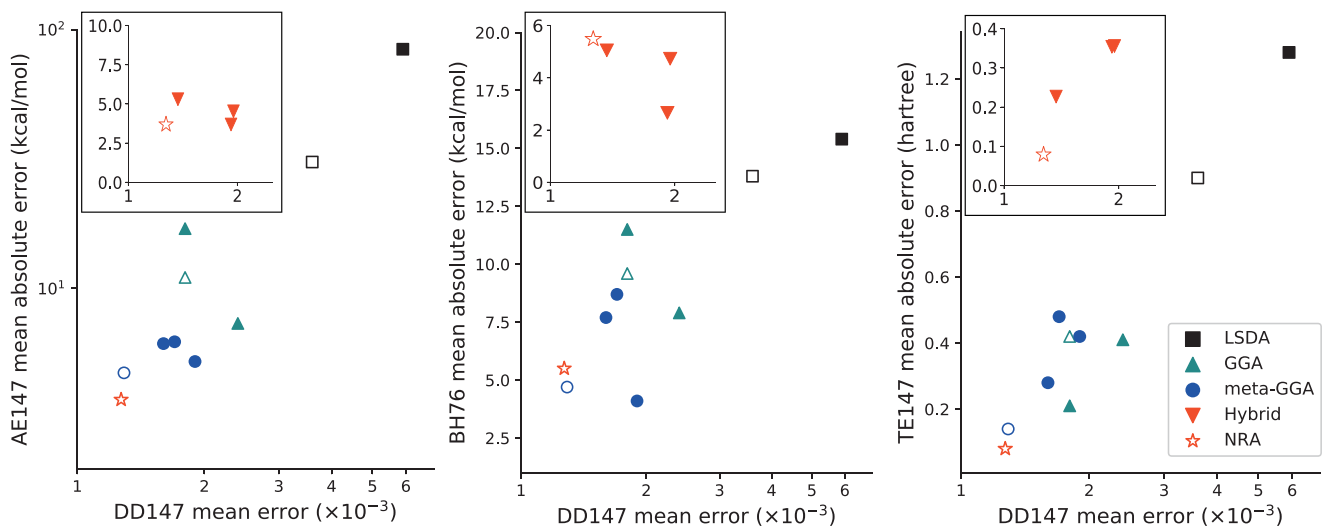}
      \caption{Improving accuracy for atomization energy (AE147), reaction barrier height (BH147), and total energy (TE147) with improving accuracy for density distribution (DD147) and increasing descriptors in the functionals, corresponding to Table 1. The closed and open markers represent accuracy of existing and the NN-based functionals respectively. }
      \label{fig:improvement}
  \end{center}
\end{figure*}

Using the trained NN-based functionals, we calculated properties of the hundreds of systems\cite{G2H0,BH76}, which consist of first to third row elements (Table 1). The performances are compared with those calculated with the existing analytic functionals. Among various functional developed to date \cite{30years}, we selected the representative (semi-)local and hybrid functionals~\cite{Slater,VWN, Becke-exchange,LYP,PBE,TPSS,SCAN,M06L, PBE0,B3LYP,M06}.  For all the properties including unreferenced ones (BH and TE), the NN-based functionals are superior or comparable in the accuracy to existing functionals at each approximation level, which are implemented in most of DFT packages and widely used.  For AE and TE, the non-local NRA type functional is comparable to the hybrid functionals, which are partly containing non-local effect and fitted to more than 100 systems.
Remarkably, the NN-based functionals are applicable to a wider range of materials and unreferenced quantities even though they are trained for the small training dataset. This fact would look non-trivial in the context of existing methods of ML for predicting material properties such as the materials informatics. This transferability reflects the advantage of our method of treating electron density, which is common to any material, as the input of ML mapping.
 It is also surprising that NN-LSDA works far better than SVWN for the test systems. Tozer et al.~\cite{NNpotential} showed that, within the LD approximation, the energy density functional cannot be determined uniquely because the xc-potential takes several values for the same local $n(r)$, as it is actually non-local. From the many choices to determine the dependence on $n(r)$, the conventional LSDA has been adjusted to uniform electron gas, while our functional can be contrasted as ``LSDA adjusted to molecular systems”. As the approximation level goes up (GGA, meta-GGA, NRA), the mapping $\mbox{\boldmath $g$}\rightarrow \varepsilon_{\rm xc}$ becomes less multivalued, hence the accuracy tend to be improved as exhibited in Fig.~\ref{fig:improvement}.  Those results suggest that systematic improvement of the functional is realized by adding further descriptors to $\mbox{\boldmath $g$}$, and by training with density distributions.

\begin{figure}[p]
  \begin{center}
      \includegraphics[clip,width=8.6cm]{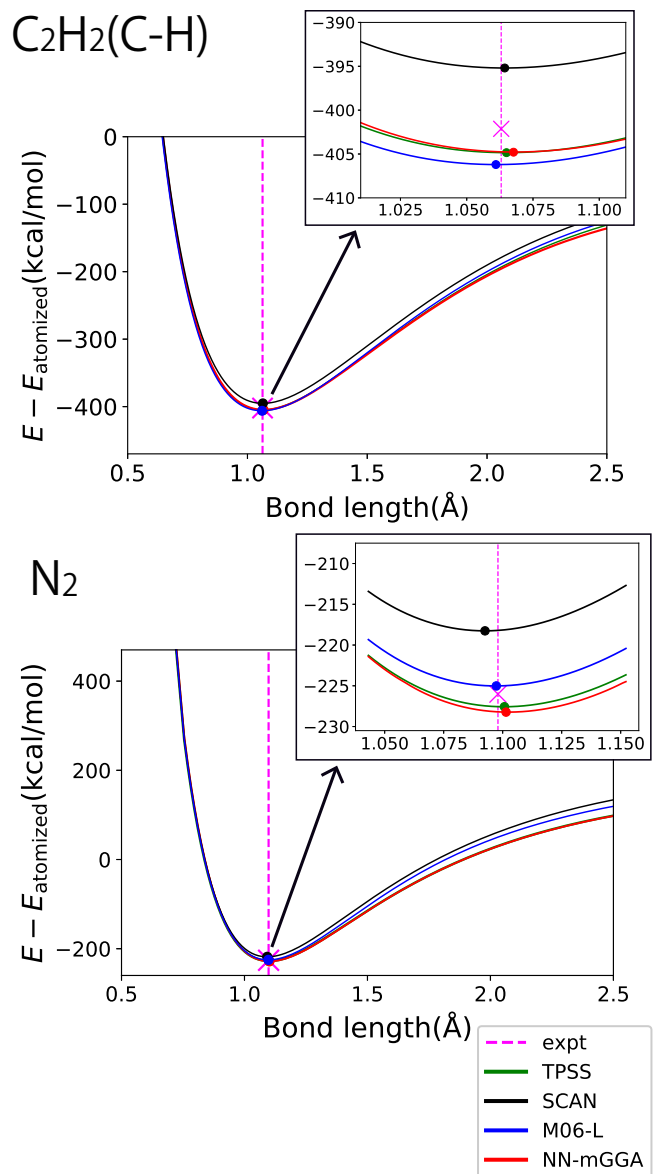}
      \caption{Dissociation curve of ${\rm C_2H_2}$ (upper) and $\rm N_2$ (lower), calculated by the NN-based and existing meta-GGA functionals. For ${\rm C_2H_2}$, the two C-H bonds are dissociated symmetrically along the original bond direction. Horizontal axis shows the bond length, and vertical axis shows energy relative to the atomized limit ($E_{\rm atomized}$). The green dashed lines and ``x'' marks show the bond lengths and the atomization energies from experiments~\cite{G2H0}. The ``o" marks show the peak of each curve. }
      \label{fig:BLcurve}
  \end{center}
\end{figure}

We also applied the NN-based meta-GGA functional to the bond dissociation of C$_2$H$_2$ and N$_2$, comparing to the existing meta-GGA functionals as shown in Fig.~\ref{fig:BLcurve}. The smoothness of their dissociation curves is not guaranteed in the conventional ML methods which skip the KS equation by relating the energy-related quantities directly with atomic configurations~\cite{crystal_graph,Chandrasekaran2019,Burke-bypassing}. This signifies the advantage of explicitly solving the KS equation, where the kinetic energy operator mitigates unphysical noises of electron density due to possible overfitting, thereby enhancing the transferability of the functional out of the training dataset, as shown in ref.\cite{NNVHxc}.

\begin{figure*}[p]
  \begin{center}
      \includegraphics[clip,width=17.8cm]{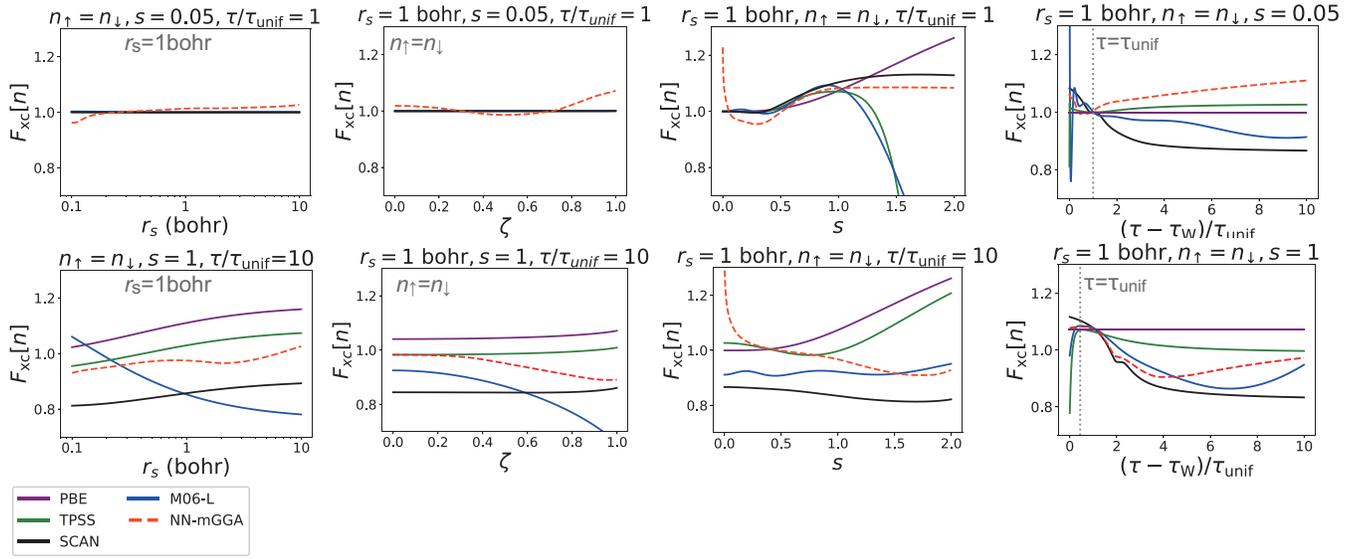}
      \caption{Behaviors of functionals around typical ranges of density distributions. The vertical axis are defined as Equation~(\ref{eq:Factor}). Meta-GGA type functionals have four variables, and the panels show the dependence on one of them while the others are fixed. $r_s\equiv (3/4\pi n)^{1/3}$ represents average distance between electrons. In a typical metal, $r_s\sim 1$bohr.  $\tau_{\rm unif}\equiv(3/10)(3\pi^2)^{2/3}n^{5/3}$, $\tau_{\rm W}\equiv|\nabla n|^2/8n$ represents $\tau$ at UEG limit and single-orbital limit respectively\cite{TPSS}. To examine the behavior in the approximately UEG limit, we set $s=0.05$ (three of the upper panels); $s>0.05$ for more than $99\%$ of the training data.}
      \label{fig:functionalshape}
  \end{center}
\end{figure*}

The NN-based meta-GGA functional is analyzed by plotting enhancement factor relative to the xc-functional of LSDA~\cite{Slater, VWN}, which corresponds to energy density in the uniform electron gas(UEG) limit: 
\begin{eqnarray}
    F_{\rm xc}[ n]\equiv \frac{\varepsilon_{\rm xc}[ n]}{\varepsilon^{\rm UEG}_{\rm xc}( n,\zeta)}.
    \label{eq:Factor}
\end{eqnarray}
As shown in Fig.~\ref{fig:functionalshape}, the NN-based functional behaves similarly to the representative functionals, except for the limits $s \rightarrow +0$ and $\tau \rightarrow \tau_{\rm W}+0$. The latter behavior is due to the fact that the training data does not cover those limits. Although the behavior of $F_{\rm xc}$ in these limits had little effect for the present test systems, it would be more desirable for broader applicability if $F_{\rm xc}$ can be constrained explicitly to satisfy several exact conditions~\cite{constraintsatis}. Such possibility has also been recently pursued in the ML functional community \cite{MLxc2,MLexact}. Interestingly, in the small $s$ regime (within the range of the training data), our $F_{\rm xc}$ showed a trend to converge to the exact asymptotic forms as functions of $r_{\rm s}$ and $\zeta$, which the other functionals are analytically enforced to satisfy, as seen in the upper left two panels. This result suggest that the proper asymptotic forms may be automatically reproduced with practical accuracy through the ML approach; this should be convenient for expanding the frontier of DFT with unconventional non-local variables such as in our NRA, for which the exact asymptotic behavior is not straightforwardly derived.

In summary, we propose a systematic ML approach to the accurate and transferable xc-functional for the KS-DFT. Our results suggest that the improvement can be made with a simple strategy: Prepare a maximally flexible NN-based functional form and train it with physical or available chemical properties of appropriate materials. Although the existing analytic functionals have been conventionally constructed through complicated heuristic procedures, the NN-based functionals can rather be constructed only with minimal assumption on the functional form.  In fact, we easily introduce the non-locality with additional descriptors such as $R$ by virtue of this ML's ability. Unlike the improvement by introducing orbital-dependent terms\cite{Jacob}, our method keeps the classical framework of solving the KS equation within the ordinary DFT computational cost to introduce the non-locality, which keeps the calculation to be executable for large systems, and can be a novel path toward the numerically exact functionals alternative to the Jacob's ladder. In this work, we have tested our method just using only three reference molecules and obtained the promising result. However, further applications remains to be studied to make a functional applicable for conventionally difficult targets such as materials with self-interaction, dispersion force or static correlation. In those non-trivial applications, we expect development of DFT over the theoretical understanding of human beings.

\section*{Methods}

\subsection*{Structure of the NN-based functional.}
We formulate the xc-energy density as
\begin{align}
\varepsilon_{\rm xc}(n,\mbox{\boldmath $g$}) 
=- n^\frac{1}{3} \frac{1}{2}\{(1+\zeta)^\frac{4}{3}
+
(1-\zeta)^\frac{4}{3}\}G_{\rm xc}^{\rm NN}(\mbox{\boldmath $g$}). 
\label{eq:energydensity}\tag{S1}
\end{align}
The first factor $ n^\frac{1}{3}$ corresponds to the Slater exchange energy density \cite{Slater}, and the second one is from spin-scaling approximation for exchange energy of uniformly-spin-polarized electron gas~\cite{PhysRevA.20.397}. They are the minimal physical conditions introduced to initial state of the NN close to the goal. The remaining correction $G_{\rm xc}^{\rm NN}$ is constructed using the fully-connected NN defined in Equation 3 with four layers:
\begin{align}
    G_{\rm xc}^{\rm NN}(\mbox{\boldmath $g$})&=1+h_4(...(h_1(\log\mbox{\boldmath $g$}))
    \label{eq:NN detailed structure}\tag{S2}
\end{align}
Note that each element of $\mbox{\boldmath $g$}$ is preprocessed as shown below if included:
\begin{align}
    n&\rightarrow \log n^\frac{1}{3},\nonumber\\
    \zeta&\rightarrow \log \left(\frac{1}{2}\{(1+\zeta)^\frac{4}{3}+(1-\zeta)^\frac{4}{3}\}\right),\nonumber\\
    s&\rightarrow \log s,\tag{S3}\\
    \tau&\rightarrow \log \left(\frac{\tau}{n^\frac{5}{3}\{(1+\zeta)^\frac{5}{3}+(1-\zeta)^\frac{5}{3}\}}\right),\nonumber\\
    R&\rightarrow\log R.\nonumber
    \label{eq:preprocess}
\end{align}
These transformations are introduced to facilitate the optimization of NN by making $\mbox{\boldmath $g$}$ dimensionless, suppressing the change in the magnitude, and regularizing the variance ranges of all input elements to be similar.
 For the activation function $f$, we adopted the smooth non-linear activation function named ``{\it Exponential~Linear~Unit}''~\cite{elu} which is defined as $f(x)={\rm max}(0,x) + {\rm min}(0,e^x-1)$. The last layer $h_H$ is designed to keep the value of $\varepsilon_{\rm xc}$ to non-positive. The dimensions of the parameter matrices and bias vectors are as follows: ${\rm dim}W_1=N\times 100,\ {\rm dim}W_2={\rm dim}W_3=100\times 100,\  {\rm dim}W_4=100\times 1,\  {\rm dim}b_1={\rm dim}b_2={\rm dim}b_3=100,\  {\rm dim}b_4=1$), where $N$ represents the number of elements in $\mbox{\boldmath $g$}$.

\subsection*{Functional with non-local density distribution.}
\label{sec:NRA}
We suggest a functional form treating non-locality by introducing a non-local descriptor:
\begin{align}
    E_{\rm xc}[ n]&= \int d{\bf r} ~n ({\bf r})\varepsilon_{\rm xc}(\mbox{\boldmath $g$}[ n] ({\bf r}))\tag{S4}\\
    \mbox{\boldmath $g$}[n] ({\bf r})&=\left(\mbox{\boldmath $g$}_{\rm local} ({\bf r}), R ({\bf r})\right)\tag{S5}\\
    R ({\bf r})&=\int d {\bf r'} n( {\bf r'})d({\bf r}, {\bf r'})\tag{S6}
    \label{eq:nlocdescript}
\end{align}
$\mbox{\boldmath $g$}_{\rm local} ({\bf r})$ represents (semi-)local descriptors such as $n({\bf r}), s({\bf r})$, or $ \tau ({\bf r})$, while $R ({\bf r})$ includes weighted density distribution around ${\bf r}$, with the weight function $d(\bf r, r')$  vanishing at the $|\bf r-r'|\rightarrow\infty$ limit. As a result of the non-locality, the functional derivative contains integration over the whole space:
\begin{align}
    V_{\rm xc}[ n] ({\bf r})
    =&
    \frac{\delta E_{\rm xc}}{\delta  n ({\bf r})}
    \nonumber\\=&
    \frac{\partial \{n ({\bf r})\varepsilon_{\rm xc}(\mbox{\boldmath $g$}[ n] ({\bf r}))\} }{\partial\mbox{\boldmath $g$}_{\rm local} ({\bf r})} 
    \cdot
    \frac{\delta\mbox{\boldmath $g$}_{\rm local} ({\bf r})}{\delta n ({\bf r})}
    +
    \int  d{\bf r'}~ \frac{\partial \{n ({\bf r'})\varepsilon_{\rm xc}(\mbox{\boldmath $g$}[ n] ({\bf r'}))\}}{\partial R ({\bf r'})} d({\bf r}, {\bf r'}).
    \nonumber\\\tag{S7}
    \label{eq:Vxcnloc}
\end{align}

We implemented those integration numerically on the same grid pints to those used in the exchange-correlation integration. The cost of evaluating the xc-potential is proportional to the square of system size.
In this work, we defined the $d({\bf r})$ as 
$\exp\left(-|\bf r- r'|/\sigma\right)$. $\sigma$ was fixed to 0.2~bohr. This is derived from the inverse of the Fermi wavenumber, which is known to be the typical distance where the contribution to exchange-correlation hole at $\bf r$ from $\bf {\bf r'}$ decays~\cite{AD}., in the H$_2$O molecule estimated from the density distribution calculated by the CCSD calculation (averaged over the grids used for the numerical integration).

\subsection*{Training the NN-based functional.}
\label{sec:MC}
We use the Monte Carlo method by repeating the following steps to train the NN:
\begin{enumerate}
    \item At the $t$-th iteration, add a perturbation $\delta {\bf w}^t$ to weights ${\bf w}^{t}$ in NN. ${\bf w}$ represents both elements in the matrices $\{W_i\}$ and the vectors $\{{\bf b}_i\}$. Each element in $\delta {\bf w}^{t}$ is generated randomly from normal distribution $N(0,\delta w)$.
    \item Conduct the KS-DFT calculation for the target molecules and atoms to evaluate the cost function $\Delta_{\rm err}^i$in Equation~(\ref{eq:ErrorFunction}) using the NN-based functional with the weight-parameters ${\bf w}^t+\delta {\bf w}^t$.
    \item  According to a random number $p$ generated from uniform distribution in (0,1) and the acceptance ratio $P$ defined as below, decide whether to accept or reject the weight-perturbation $\delta {\bf w}^t$.
    \begin{align}
        P = \exp\left(-\frac{\Delta_{\rm err}^t-\Delta_{\rm err}^{t-1}}{T\Delta_{\rm err}^{t-1}}\right)\tag{S8}
    \end{align}
    If $P>1$: Set ${\bf w}^{t+1}={\bf w}^t+\delta {\bf w}$ and restart from step 2. \\
     If $p<P<1$: Set ${\bf w}^{t+1}={\bf w}^t+\delta {\bf w}$ and restart from step 1. \\
    If $P<p$: Set ${\bf w}^{t+1}={\bf w}^t$ and restart from step 1.
\end{enumerate}

We repeated those steps with making $\delta w$ and $T$ smaller, until the error function becomes small enough. 
The cost function $\Delta_{\rm err}$ is defined as
\begin{align}
\Delta_{\rm err}=
    &c_1(\Delta^{\rm G2} {\rm AE}_{\rm H_2O}+\Delta^{\rm G2} {\rm AE}_{\rm NH_3}+\Delta^{\rm G2} {\rm AE}_{\rm NO})/E_0\nonumber\\
    &+c_2(\Delta^{\rm CCSD}n_{\rm H_2O}+\Delta n^{\rm CCSD}_{\rm NH_3}+\Delta^{\rm CCSD} n_{\rm NO}),\nonumber\\\tag{S9}
    \label{eq:ErrorFunction}
\end{align}
where $\Delta^{\rm G2} {\rm AE}$ represents the absolute deviation of the atomization energy in hartree from the G2 calculation, and $E_0$ was set to 1 hartree.
The $\Delta^{\rm CCSD} n$ represents the error between ${n}$ obtained by the DFT and the CCSD calculation:
\begin{align}
    \Delta^{\rm CCSD} n_{\rm M}=\frac{1}{N_e}\sqrt{\int d{\bf r}~\left( n^{\rm DFT}_{\rm M}({\bf r})- n^{\rm CCSD}_{\rm M}({\bf r})\right)^2},\tag{S10}
    \label{eq:densityerror}
\end{align}
where $N_e$ represents the number of electrons in the molecule $M$.
The integrations are conducted numerically on grid points which is the same to those used in exchange-correlation integration of KS equation (See PySCF ~\cite{PySCF} document for details, the default DFT setting is adopted in all calculation through this work.). The $E_0$ is adjusted so that the contributions from the two terms become almostly same at the initial step of the training. $c_2/c_1$ determines the balance of the two terms. In this study, $E_0$ and $c_2/c_1$ was fixed to 10 hartree and 1 respectively, for the training of any type of functional. 

In the training of each NN-based functional, the whole steps were conducted for about 300 times. The initial $T$ and $\delta W$ are set to 0.1 and 0.01. At the final step, they are reduced to 0.06 and 0.005 respectively. The whole steps were conducted in parallel with 160 threads by ISSP System C, and the weight parameters which minimize the cost function the most were adopted finally. \\\\

\subsection*{Computational details.} All of the DFT and the CCSD calculation in our work were implemented using PySCF code with 6-311++G(3df,3pd) basis set was used for both in training of the NN-based functionals and in testing accuracies of the functionals. For the NN implementation, we used Pytorch package~\cite{Pytorch}.  The NN-based functionals could cause a convergence issue due to poor extrapolation when they are applied to density far from that included in training dataset, therefore the initial density guess of self-consistent DFT should be close enough to the final destination. In this work, initial guess of Kohn-Sham density is given by a superposition of atomic density, and under this condition, there was no convergence issues. 

\subsection*{Data availability}
The trained NN parameters are available in {\it https://github.com/ml-electron-project/NNfunctional} with usages in the PySCF code.

\section*{Acknowledgements}

R.N. thanks Takahito Nakajima for enlightening comments. Part of the calculations were performed at the Supercomputer Center at the Institute for Solid State Physics in the University of Tokyo.

\section*{Author contributions}

R.N. established the method and conducted the calculation.
R.N., R.A., and O.S. contributed in analyzing the results and writing the manuscript.

\section*{Competing interests}
The authors declare no competing interests.

\newpage

\newpage

\newpage

\newpage

\begin{table}[p]
\centering
  \begin{tabular}{c c c c c c} \hline \hline
      & ${\rm AE147}^{\rm a}$  & ${\rm DD147}^{\rm b}$  &BH76$^{\rm c}$& TE147$^{\rm d}$  \\ 
      & (kcal/mol)  &- & (kcal/mol)&(hartree) \\ 
      \hline
      SVWN & 84.2 & 0.0059&15.4  &1.28\\ \vspace{1mm}
      \textsl{NN-LSDA} & 30.9 &0.0036& 13.8& 0.90\\ 
      BLYP & 7.3 &0.0024& 7.9 &0.41 \\ 
      PBE  & 17.0&0.0018& 11.5 &0.21\\ \vspace{1mm}
      \textsl{NN-GGA} & 11.0 &0.0018& 9.6 &0.42\\ 
       TPSS  & 6.2&0.0017& 8.7 &0.48\\ 
      SCAN &  6.1&0.0016& 7.7   &0.28\\  
      M06-L  & 5.2&0.0019& 4.1 &0.42\\ \vspace{1mm}
      \textsl{NN-metaGGA}  & 4.7&0.0013& 4.7  &0.14\\
      PBE0  & 5.3&0.0014& 5.0  & 0.23\\
      B3LYP  & 4.5&0.0019& 4.7  & 0.36\\ \vspace{1mm}
      M06  & 3.7&0.0019& 2.7  & 0.35\\ 
      \textsl{NN-NRA}  & 3.7&0.0013& 5.5  & 0.08\\ \hline \hline
  \end{tabular}
    \caption{Mean absolute error (MAE) of NN-based functionals and existing analytic functionals for (a)atomization energy (AE), (b)density distribution (DD, see Equation~(S10) in the Methods for the definition), (c)chemical barrier height(BH), and (d)total energy (TE). (a) and (d) refer to the G2 method for 147 molecules listed in~\cite{G2H0}, (b) to the CCSD~\cite{CCSD} calculation for the same molecules, and (c) to the 76 reactions in~\cite{BH76}. The prefix ``NN'' represents the NN-based functional. All of the DFT and CCSD calculation were performed using the 6-311++G(3df,3pd) basis set and implemented in PySCF\cite{PySCF}, except for the DFT with existing LSDA, GGA, meta-GGA functionals for BH, where the value is cited from~\cite{SCAN}, because of ill-convergence problems of some functionals in PySCF for systems including H$_2$. For the same reason, the H$_2$ molecule listed in~\cite{G2H0} is excluded from the MAE in (a), (c), and (d). }
  \label{tab:accuracy}
\end{table}

\end{document}